\def\year{2021}\relax
\documentclass[letterpaper]{article}
\usepackage{aaai21} 
\usepackage{times} 
\usepackage{helvet} 
\usepackage{courier} 
\usepackage[hyphens]{url} 
\usepackage{graphicx} 
\usepackage{graphicx} 
\usepackage{natbib}
\usepackage{caption}
\usepackage{amsmath,amssymb,amsfonts}
\DeclareMathOperator*{\argmax}{arg\,max}

\usepackage{multirow}
\usepackage{enumitem}
\usepackage{caption}
\usepackage{amsmath,amssymb,amsfonts}
\usepackage{algorithmic}
\usepackage{graphicx}
\usepackage{textcomp}
\usepackage{xcolor}
\usepackage{comment}

\newcommand{\student}{{OPD\textsuperscript{S}}}
\newcommand{\studentg}{{OPD}}
\newcommand{\teacher}{{OPD\textsuperscript{T}}}

\newcommand{\bs}{\boldsymbol}
\newcommand{\s}{\bs{s}}
\newcommand{\bphi}{\bs{\phi}} 
\newcommand{\btheta}{\bs{\theta}} 
\newcommand{\minisection}[1]{\vspace{5pt}\noindent\textbf{#1.}}
\newcommand{\tabincell}[2]{\begin{tabular}{@{}#1@{}}#2\end{tabular}}

\begin{document}
%
\title{Universal Trading for Order Execution with Oracle Policy Distillation}
\author {
    Yuchen Fang, \textsuperscript{\rm 1}\thanks{The work was conducted when Yuchen Fang was doing internship at Microsoft Research Asia.}
    Kan Ren, \textsuperscript{\rm 2}
    Weiqing Liu, \textsuperscript{\rm 2}
    Dong Zhou, \textsuperscript{\rm 2} \\
    Weinan Zhang, \textsuperscript{\rm 1}
    Jiang Bian, \textsuperscript{\rm 2}
    Yong Yu, \textsuperscript{\rm 1}
    Tie-Yan Liu\textsuperscript{\rm 2}\\
}
\affiliations {
    \textsuperscript{\rm 1}Shanghai Jiao Tong University,
    \textsuperscript{\rm 2}Microsoft Research \\
    \{arthur\_fyc, wnzhang\}@sjtu.edu.cn, yyu@apex.sjtu.edu.cn\\
    \{kan.ren, weiqing.liu, zhou.dong, jiang.bian, tie-yan.liu\}@microsoft.com
}
\maketitle
\begin{abstract}
As a fundamental problem in algorithmic trading, order execution aims at fulfilling a specific trading order, either liquidation or acquirement, for a given instrument. Towards effective execution strategy, recent years have witnessed the shift from the analytical view with model-based market assumptions to model-free perspective, i.e., reinforcement learning, due to its nature of sequential decision optimization. However, the noisy and yet imperfect market information that can be leveraged by the policy has made it quite challenging to build up sample efficient reinforcement learning methods to achieve effective order execution. In this paper, we propose a novel universal trading policy optimization framework to bridge the gap between the noisy yet imperfect market states and the optimal action sequences for order execution. Particularly, this framework leverages a policy distillation method that can better guide the learning of the common policy towards practically optimal execution by an oracle teacher with perfect information to approximate the optimal trading strategy. The extensive experiments have shown significant improvements of our method over various strong baselines, with reasonable trading actions.

\end{abstract}

\section{Introduction}
\begin{figure}[ht]
	\begin{centering}
		\includegraphics[width=0.65\columnwidth]{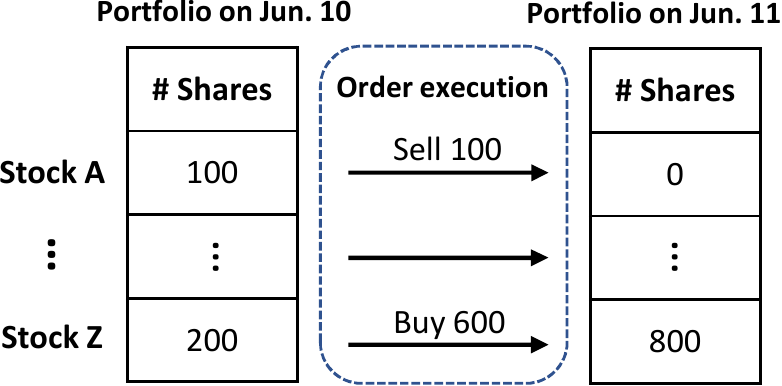}
		\caption{An example of portfolio adjustment which requires order execution.}
		\label{fig:pos-adjust} 
		\vspace{-15pt}
	\end{centering}
\end{figure}

Financial investment, aiming to pursue long-term maximized profits, is usually behaved in the form of a sequential process of continuously adjusting the asset portfolio.
One indispensable link of this process is \textit{order execution}, consisting of the actions to adjust the portfolio.
Take stock investment as an example, the investors in this market construct their portfolios of a variety of instruments.  As illustrated in Figure~\ref{fig:pos-adjust}, to adjust the position of the held instruments, the investor needs to sell (or buy) a number of shares through executing an order of liquidation (or acquirement) for different instruments.
Essentially, the goal of order execution is two-fold: it does not only requires to fulfill the whole order but also targets a more economical execution with maximizing profit gain (or minimizing capital loss).

As discussed in \cite{cartea2015algorithmic}, the main challenge of order execution lies in a trade-off between avoiding harmful ``\textit{market impact}'' caused by large transactions in a short period and restraining ``\textit{price risk}'', i.e., missing good trading opportunities, due to slow execution.
Previously, traditional analytical solutions often adopt some stringent assumptions of market liquidity, i.e., price movements and volume variance, then derive some closed-form trading strategies based on stochastic control theory with dynamic programming principle \cite{bertsimas1998optimal,almgren2001optimal,cartea2015optimal,cartea2016incorporating,cartea2015algorithmic}. 
Nonetheless, these model-based\footnote{Here `model' corresponds to some market price assumptions, to distinguish the environment model in reinforcement learning.} methods are not likely to be practically effective because of the inconsistency between market assumption and reality.

According to the order execution's trait of sequential decision making, reinforcement learning (RL) solutions \cite{nevmyvaka2006reinforcement,ning2018double,lin2020end} have been proposed from a model-free perspective and been increasingly leveraged to optimize execution strategy through interacting with the market environment purely based on the market data.
As data-driven methods, RL solutions are not necessarily confined by those unpractical financial assumptions and are thus more capable of capturing the market's microstructure for better execution.

Nevertheless, RL solutions may suffer from a vital issue, i.e., the largely noisy and imperfect information.
For one thing, the noisy data \cite{wu2020conditional} could lead to a quite low sample efficiency of RL methods and thus results in poor effectiveness of the learned order execution policy.
More importantly, the only information that can be utilized when taking actions is the historical market information, without any obvious clues or accurate forecasts of the future trends of the market price\footnote{We define `market price' as the averaged transaction price of the whole market at one time which has been widely used in literature \cite{nevmyvaka2006reinforcement,ning2018double}.} or trading activities.
Such issues indeed place a massive barrier for obtaining a policy to achieve the optimal trading strategy with profitable and reasonable trading actions.

\begin{figure}[h]
	\begin{centering}
		\includegraphics[width=0.55\columnwidth]{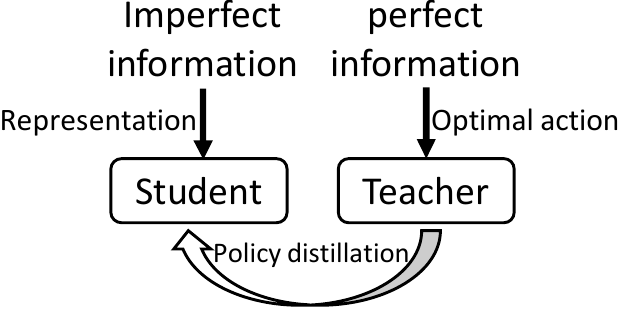}
		\caption{Brief illustration of the oracle policy distillation.}
		\label{fig:policy-dist} 
	\end{centering}
\end{figure}

To break this barrier, in this paper, we propose a universal policy optimization framework for order execution.
Specifically, this framework introduces a novel policy distillation approach in order for bridging the gap between the noisy yet imperfect market information and the optimal order execution policy. 
More concretely, as shown in Figure~\ref{fig:policy-dist}, this approach is essentially a teacher-student learning paradigm, in which the teacher, with access to perfect information, is trained to be an oracle to figure out the optimal trading strategy, and the student is learned by imitating the teacher's optimal behavior patterns.
For practical usage, only the common student policy with imperfect information, would be utilized \textit{without} teacher or any future information leakage.
Also note that, since our RL-based model can be learned using all the data from various instruments, it can largely alleviate the problem of model over-fitting.

The main contributions of our paper include:
\begin{itemize}
    \item We show the power of learning-based yet model-free method for optimal execution, which not only surpasses the traditional methods, but also illustrates reasonable and profitable trading actions.
    \item The teacher-student learning paradigm for policy distillation may beneficially motivate the community, to alleviate the problem between imperfect information and the optimal decision making.
    \item To the best of our knowledge, it is the first paper exploring to learn from the data of various instruments, and derive a universal trading strategy for optimal execution. 
\end{itemize}

\section{Related Works}
\subsection{Order Execution}
Optimal order execution is originally proposed in \cite{bertsimas1998optimal} and the main stream of the solutions are model-based.
\cite{bertsimas1998optimal} assumed a market model where the market price follows an arithmetic random walk in the presence of linear price impact function of the trading behavior.
Later in a seminal work \cite{almgren2001optimal}, the Almgren-Chriss model has been proposed which extended the above solution and incorporated both temporary and permanent price impact functions with market price following a Brownian motion process.
Both of the above methods tackled order execution as a financial model and solve it through stochastic control theory with dynamic programming principle.

Several extensions to these pioneering works have been proposed in the past decades \cite{cartea2016incorporating,gueant2015general,gueant2016financial,casgrain2019trading} with modeling of a variety of market features.
From the mentioned literature, several closed-form solutions have been derived based on stringent market assumptions. 
However, they might not be practical in real-world situations thus it appeals to non-parametric solutions \cite{ning2018double}.

\subsection{Reinforcement Learning}
RL attempts to optimize an accumulative numerical reward signal by directly interacting with the environment \cite{sutton2018reinforcement} under few model assumptions such as Markov Decision Process (MDP).
RL methods have already shown superhuman abilities in many applications, such as game playing \cite{mnih2015human,li2020suphx,silver2016mastering}, resource dispatching \cite{li2019cooperative,tang2019deep}, financial trading \cite{moody1997optimization,moody1998reinforcement,moody2001learning} and portfolio management \cite{jiang2017cryptocurrency,ye2020reinforcement}.

Besides the analytical methods mentioned above, RL gives another view for sequential decision optimization in order execution problem.
Some RL solutions \cite{hendricks2014reinforcement,hu2016optimal,daberius2019deep} solely extend the model-based assumptions mentioned above to either evaluate how RL algorithm might improve over the analytical solutions \cite{daberius2019deep,hendricks2014reinforcement}, or test whether the MDP is still viable under the imposed market assumptions \cite{hu2016optimal}.
However, these RL-based methods still rely on financial model assumption of the market dynamics thus lack of practical value.

Another stream of RL methods abandon these financial market assumptions and utilize model-free RL to optimize execution strategy \cite{nevmyvaka2006reinforcement,ning2018double,lin2020end}.
Nevertheless, these works faces challenges of utilizing imperfect and noisy market information.
And they even train separate strategies for each of several manually chosen instruments, which is not efficient and may result in over-fitting.
We propose to handle the issue of market information utilization for optimal policy optimization, while training a universal strategy for all instruments with general pattern mining.

\subsection{Policy Distillation}
This work is also related to policy distillation \cite{rusu2015policy}, which incorporates knowledge distillation \cite{hinton2015distilling} in RL policy training and has attracted many researches studying this problem \cite{parisotto2015actor,teh2017distral,yin2017knowledge,green2019distillation,lai2020dual}.
However, these works mainly focus on multi-task learning or model compression, which aims at deriving a comprehensive or tiny policy to behave normally in different game environments.
Our work dedicates to bridge the gap between the imperfect environment states and the optimal action sequences through policy distillation, which has not been properly studied before.

\section{Methodology}
In this section, we first formulate the problem and present the assumptions of our method including MDP settings and market impact.
Then we introduce our policy distillation framework and the corresponding policy optimization algorithm in details.
Without loss of generality, we take \textit{liquidation}, i.e., to sell a specific number of shares, as an on-the-fly example in the paper, and the solution to acquirement can be derived in the same way.
\subsection{Formulation of Order Execution}
Generally, order execution is formulated under the scenario of trading within a predefined time horizon, e.g., one hour or a day.
We follow a widely applied simplification to the reality~\cite{cartea2015algorithmic,ning2018double} by which trading is based on discrete time space.

Under this configuration, we assume there are $T$ timesteps $\{0, 1, \ldots, T-1\}$, each of which is associated with a respective price for trading $\{p_0, p_1, \ldots, p_{T-1}\}$.
At each timestep $t \in \{0, 1, \ldots, T-1\}$, the trader will 
propose to trade a volume of $q_{t+1} \geq 0$ shares, the trading order of which will then be actually executed with the execution price $p_{t+1}$, i.e., the market price at the timestep $(t+1)$.
With the goal of maximizing the revenue with completed liquidation, the objective of optimal order execution, assuming totally $Q$ shares to be liquidated during the whole time horizon, can be formulated as
\begin{equation}\label{eq:obj}
\begin{aligned}
    \mathop{\arg \max}_{q_1,q_2,\ldots, q_T}  \sum_{t=0}^{T-1} (q_{t+1} \cdot p_{t+1}), ~~ 
    s.t. \sum_{t=0}^{T-1} q_{t+1} = Q ~.
\end{aligned}
\end{equation}
The average execution price (AEP) is calculated as
$
    \bar{P} = \frac{\sum_{t=0}^{T-1} (q_{t+1} \cdot p_{t+1})}{\sum_{t=0}^{T-1} q_{t+1}} = \sum_{t=0}^{T-1} \frac{q_{t+1}}{Q}\cdot p_{t+1} ~.
$
The equations above expressed that, for liquidation, the trader needs to maximize the average execution price $\bar{P}$ so that to gain as more revenue as possible.
Note also that, for acquirement, the objective is reversed and the trader is to minimize the execution price to reduce the acquirement cost.

\subsection{Order Execution as a Markov Decision Process}\label{sec:mdp}
Given the trait of sequential decision-making, order execution can be defined as a Markov Decision Process. To solve this problem, RL solutions are used to learn a policy $\pi \in \Pi$ to control the whole trading process through proposing an action $a \in \mathcal{A}$ given the observed state $\s \in \mathcal{S}$, while maximizing the total expected rewards based on the reward function $R(\s, a)$.

\minisection{State} The observed state $\s_t$ at the timestep $t$ describes the status information of the whole system including both the trader and the market variables. 
Two types of widely-used information~\cite{nevmyvaka2006reinforcement,lin2020end} for the trading agent are private variable of the trader and public variable of the market.
Private variable includes the elapsed time $t$ and the remained inventory $(Q-\sum_{i=1}^t q_i)$ to be executed.
As for the public variable, it is a bit tricky since the trader only observes \textit{imperfect} market information.
Specifically, one can only take the past market history, which have been observed \textit{at or before} the time $t$, into consideration to make the trading decision for the next timestep.
The market information includes open, high, low, close, average price and transaction volume of each timestep.

\minisection{Action} Given the observed state $\s_t$, the agent proposes the decision $a_t = \pi(\s_t)$ according to its trading policy $\pi$, where $a_t$ is discrete and corresponds to the proportion of the target order $Q$.
Thus, the trading volume to be executed at the next time can be easily derived as $q_{t+1} = a_t \cdot Q$, and each action $a_t$ is the standardized trading volume for the trader.

According to Eq.~(\ref{eq:obj}), we follow the prior works \cite{ning2018double,lin2020end} and assume that the summation of all the agent actions during the whole time horizon satisfies the constraint of order fulfillment that $\sum_{t=0}^{T-1} a_t = 100\%$.
So that the agent will have to propose $a_{T-1} = \max\{ 1-\sum_{i=0}^{T-2}a_i, \pi(\s_{T-1}) \}$ to fulfill the order target, at the last decision making time of $(T-1)$.
However, leaving too much volume to trade at the last trading timestep is of high price risk, which requires the trading agent to consider future market dynamics at each time for better execution performance.

\minisection{Reward}
The reward of order execution in fact consists of two practically conflicting aspects, trading profitability and market impact penalty.
As shown in Eq.~(\ref{eq:obj}), we can partition and define an immediate reward after decision making at each time.

To reflect the trading profitability caused by the action, we formulate a positive part of the reward as volume weighted \textit{price advantage}:
\begin{equation}\label{eq:rew}
    \hat{R}_t^{+}(\s_t, a_t)
    = \frac{q_{t+1}}{Q} \cdot \overbrace{ \left( \frac{p_{t+1} - \Tilde{p} }{ \Tilde{p}} \right) }^{ \text{price normalization} }
    = a_t \left( \frac{p_{t+1}}{\Tilde{p}} - 1 \right) ~,
\end{equation}
where $\Tilde{p}=\frac{1}{T} \sum_{i=0}^{T-1} p_{i+1}$ is the averaged original market price of the whole time horizon.

Here we apply normalization onto the absolute revenue gain in Eq.~(\ref{eq:obj}) to eliminate the variance of different instrument prices and target order volume, to optimize a universal trading strategy for various instruments.

Note that although we utilize the future price of the instrument to calculate $\hat{R}^+$, the reward is not included in the state thus would not influence the actions of our agent or cause any information leakage. 
It would only take effect in back-propagation during training.
As $\Tilde{p}$ varies within different episodes, this MDP might be non-stationary. 
There are some works on solving this problem \cite{gaon2020reinforcement}, however this is not the main focus of this paper and would not influence the conclusion.

To account for the potential that the trading activity may have some impacts onto the current market status, following \cite{ning2018double,lin2020end}, we also incorporate a quadratic penalty 
\begin{equation}
    \hat{R}_t^{-} = -\alpha (a_t)^2
\end{equation}
on the number of shares, i.e., proportion $a_t$ of the target order to trade at each decision time.
$\alpha$ controls the impact degree of the trading activities.
Such that the final reward is 
\begin{equation}\label{eq:instant-reward}
\begin{aligned}
    R_t(\s_t, a_t) 
    &= \hat{R}_t^{+}(\s_t, a_t) + \hat{R}_t^{-}(\s_t, a_t) \\
    &= \left( \frac{p_{t+1}}{\Tilde{p}} - 1 \right) a_t - \alpha \left( a_t \right)^2 ~.
\end{aligned}
\end{equation}

The overall value, i.e., expected cumulative discounted rewards of execution for each order following policy $\pi$, is $\mathbb{E}_\pi[ \sum_{t=0}^{T-1} \gamma^t R_t ]$ and the final goal of reinforcement learning is to solve the optimization problem as
\begin{equation}\label{eq:exp-rewards}
    \mathop{\arg \max}_{\pi} 
    \mathbb{E}_\pi \Big[ \sum_{t=0}^{T-1} \gamma^t R_t(\s_t, a_t) \Big] ~.
\end{equation}

As for the numerical details of the market variable and the corresponding MDP settings, please refer to the supplementary including the released code.

\subsubsection{Assumptions} Note that there are two main assumptions adopted in this paper.
Similar to \cite{lin2020end},
(i) the temporary market impact has been adopted as a reward penalty and we assume that the market is resilient and will bounce back to the equilibrium at the next timestep.
(ii) We either ignore the commissions and exchange fees as the these expense is relatively small fractions for the institutional investors that we are mainly aimed at.
\begin{figure}
  \begin{center}
   \includegraphics[width=0.65\columnwidth]{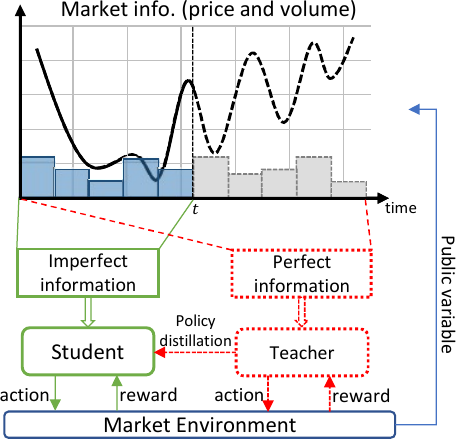}
  \end{center}
  \caption{The framework of oracle policy distillation. All the modules with dotted lines, i.e. teacher and policy distillation procedure, are only used during the \textit{training} phase, and would be removed during the \textit{test} phase or in practice.}\label{fig:framework}
  \vspace{-10pt}
\end{figure}

\subsection{Policy Distillation and Optimization}
In this section, we explain the underlying intuition behind our policy distillation and illustrate the detailed optimization framework.

\subsubsection{Policy Distillation}
As aforementioned, our goal is to bridge the gap between the imperfect information
and the optimal trading action sequences.
An end-to-end trained RL policy may not effectively capture the representative patterns from the imperfect yet noisy market information.
Thus it may result in much lower sample efficiency, especially for RL algorithms.
To tackle with this, we propose a two stage learning paradigm with teacher-student policy distillation.

We first explain the imperfect and perfect information.
In additional to the private variable including left time and the left unexecuted order volume, at time $t$ of one episode, the agent will also receive the state of the public variable, i.e., the market information of the specific instrument price and the overall transaction volume within the whole market.
However, the actual trader as a common agent only has the history market information which is collected before the decision making time $t$.
We define the history market information as \textit{imperfect} information and the state with that as the common state $\s_t$.
On the contrary, assuming that one oracle has the clue of the future market trends, she may derive the optimal decision sequence with this \textit{perfect} information, as notated as $\Tilde{\s}_t$.

As illustrated in Figure~\ref{fig:framework}, we incorporate a teacher-student learning paradigm.
\begin{itemize}
    \item \textbf{Teacher} plays a role as an oracle whose goal is to achieve the optimal trading policy $\Tilde{\pi}_{\bphi}(\cdot | \Tilde{\s}_t)$ through interacting with the environment given the perfect information $\Tilde{\s}_t$, where $\bphi$ is the parameter of the teacher policy.
    \item \textbf{Student} itself learns by interacting with the environment to optimize a common policy $\pi_{\btheta}(\cdot | \s_t)$ with the parameter $\btheta$ given the imperfect information $\s_t$.
\end{itemize}

To build a connection between the imperfect information to the optimal trading strategy, we implement a policy distillation loss $L_d$ for the student.
A proper form is the negative log-likelihood loss measuring how well the student's decision matching teacher's action as
\begin{equation}
    L_d = - \mathbb{E}_t \left[ \log \text{Pr} ( a_t = \Tilde{a}_t | \pi_{\btheta}, \s_t; \pi_{\bphi}, \Tilde{\s}_t ) \right] ~,
\end{equation}
where $a_t$ and $\Tilde{a}_t$ are the actions taken from the policy $\pi_{\btheta}(\cdot | \s_t)$ of student and $\pi_{\bphi}(\cdot | \Tilde{\s}_t)$ of teacher, respectively.
$\mathbb{E}_t$ denotes the empirical expectation over timesteps.

Note that, the teacher is utilized to achieve the optimal action sequences which will then be used to guide the learning of the common policy (student). 
Although, with perfect information, it is possible to find the optimal action sequence using a searching-based method, this may require human knowledge and is quite inefficient with extremely high computational complexity as analyzed in our supplementary.
Moreover, another key advantage of the learning-based teacher lies in the universality, enabling to transfer this method to the other tasks, especially when it is very difficult to either build experts or leverage human knowledge.

\subsubsection{Policy Optimization}
Now we move to the learning algorithm of both teacher and student.
Each policy of teacher and student is optimized separately using the same algorithm, thus we describe the optimization algorithm with the student notations, and that of teacher can be similarly derived by exchanging the state and policy notations.

With the MDP formulation described above, we utilize Proximal Policy Optimization (PPO) algorithm \cite{schulman2017proximal} in actor-critic style to optimize a policy for directly maximizing the expected reward achieved in an episode. 
PPO is an on-policy RL algorithm which seeks towards the optimal policy within the trust region by minimizing the objective function of policy as 
\begin{equation}\label{eq:policy-loss}
\small{
    L_p(\btheta) = -\mathbb{E}_t\left[\frac{\pi_{\btheta}(a_t | \s_t)}{\pi_{\btheta_{\text{old}}}(a_t | \s_t)}\hat{A}(\s_t, a_t) - \beta \text{KL}\left[\pi_{\btheta_{\text{old}}}(\cdot | \s_t), \pi_{\btheta}(\cdot | \s_t)\right]\right] .
    }
\end{equation}
Here $\btheta$ is the current parameter of the policy network, $\btheta_{\text{old}}$ is the previous parameter before the update operation.
$\hat{A}(\s_t, a_t)$ is the estimated advantage calculated as 
\begin{equation}
\hat{A}(\s_t, a_t) = R_{t}(\s_t, a_t) + \gamma V_{\btheta}(\s_{t+1}) - V_{\btheta}(\s_t) ~.
\end{equation}
It is a little different to that in \cite{schulman2017proximal} which utilizes generalized advantage estimator (GAE) for advantage estimation as our time horizon is relatively short and GAE does not bring better performance in our experiments.
$V_{\btheta}(\cdot)$ is the value function approximated by the critic network, which is optimized through a value function loss 
\begin{equation}\label{eq:ppo}
    L_v(\btheta) = \mathbb{E}_t\left[\|V_{\btheta}(\s_t) - V_t\|_2\right] ~.
\end{equation}
$V_t$ is the empirical value of cumulative future rewards calculated as below
\begin{equation}
    V_t = \sum_{t' = t}^{T-1} \mathbb{E}\left[\gamma^{T-t'-1}R_{t'}(s_{t'}, a_{t'})\right] ~.
\end{equation}
The penalty term of KL-divergence in Eq.~(\ref{eq:policy-loss}) controls the change within an update iteration, whose goal is to stabilize the parameter update.
And the adaptive penalty parameter $\beta$ is tuned according to the result of KL-divergence term following \cite{schulman2017proximal}.

\begin{figure}
  \begin{center}
   \includegraphics[width=0.5\columnwidth]{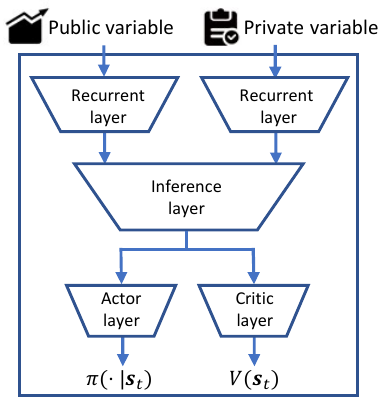}
  \end{center}
  \caption{The overall structure of the policy network.}\label{fig:policy}
  \vspace{-10pt}
\end{figure}

As a result, the overall objective function 
of the student includes the policy loss $L_p$, the value function loss $L_v$ and the policy distillation loss $L_d$ as
\begin{equation}\small{
    \begin{aligned}
    L(\btheta) &= \overbrace{ L_p + \lambda L_v }^{\text{policy optimization}} + \overbrace{ \mu L_d }^{\text{policy distillation}} ~,\\
    \end{aligned}
    }
\end{equation}
where $\lambda$ and $\mu$ are the weights of value function loss and the distillation loss.
The policy network is optimized to minimize the overall loss with the gradient descent method.
Please also be noted that the overall loss function of the teacher $L(\bphi)$ does not have the policy distillation loss.

\subsubsection{Policy Network Structure}
As is shown in Figure~\ref{fig:policy}, there are several components in the policy network where the parameter $\btheta$ is shared by both actor and critic.
Two recurrent neural networks (RNNs) have been used to conduct high-level representations from the public and private variables, separately.
The reason to use RNN is to capture the temporal patterns for the sequential evolving states within an episode.
Then the two obtained representations are combined and fed into an inference layer to calculate a comprehensive representation for the actor layer and and the critic layer.
The actor layer will propose the final action distribution under the current state.
For training, following the original work of PPO \cite{schulman2017proximal}, the final action are sampled w.r.t. the produced action distribution $\pi(\cdot|\s_t)$.
As for evaluation and testing, on the contrary, the policy migrates to fully exploitation thus the actions are directly taken by $a_t=\argmax_a \pi(\cdot|\s_t)$ without exploration.
The detailed implementations are described in supplemental materials.

\subsubsection{Discussion}
Compared to previous related studies~\cite{ning2018double,lin2020end},
the main novelty of our method lies in that we incorporate a novel policy distillation paradigm to help common policy to approximate the optimal trading strategy more effectively.
In addition, other than using a different policy network architecture, we also proposed a normalized reward function for universal trading for all instruments which is more efficient than training over single instrument separately.
Moreover, we find that the general patterns learned from the other instrument data can improve trading performance and we put the experiment and discussions in the supplementary.

\section{Experiments}
In this section, we present the details of the experiments.
The implementation descriptions and codes are in the supplementary in this link\footnote{https://seqml.github.io/opd/}.

We first raise three research questions (RQs) to lead our discussion in this section. 
\textbf{RQ1}: Does our method succeed in finding a proper order execution strategy which applies universally on all instruments?
\textbf{RQ2}: Does the oracle policy distillation method help our agent to improve the overall trading performance? 
\textbf{RQ3}: What typical execution patterns does each compared method illustrate?

\subsection{Experiment settings}

\subsubsection{Compared methods}
We compare our method and its variants with the following baseline order execution methods.
\begin{itemize}[align=left]
    \item[\textbf{TWAP}] (Time-weighted Average Price) is a model-based strategy which equally splits the order into $T$ pieces and evenly execute the same amount of shares at each timestep. It has been proven to be the optimal strategy under the assumption that the market price follows Brownian motion \cite{bertsimas1998optimal}.
    \item [\textbf{AC}] (Almgren-Chriss) is a model-based method \cite{almgren2001optimal}, which analytically finds the efficient frontier of optimal execution. 
    We only consider the temporary price impact for fair comparison.
    \item [\textbf{VWAP}] (Volume-weighted Average Price) is another model-based strategy which distributes orders in proportion to the (empirically estimated) market transaction volume in order to keep the execution price closely tracking the market average price ground truth \cite{kakade2004competitive,bialkowski2008improving}.
    \item [\textbf{DDQN}] (Double Deep Q-network) is a value-based RL method \cite{ning2018double} and adopts state engineering optimizing for individual instruments.
    \item [\textbf{PPO}] is a policy-based RL method \cite{lin2020end} which utilizes PPO algorithm with a sparse reward to train an agent with a recurrent neural network for state feature extraction. The reward function and the network architecture are different to our method.
    \item[\textbf{OPD}] is our proposed method described above, which has two other variants for ablation study: \textbf{\student}~is the pure student trained \textit{without} oracle guidance and \textbf{\teacher}~is the teacher trained with perfect market information.
\end{itemize}

Note that, for fair comparison, all the learning-based methods, i.e., DDQN, PPO and OPD, have been trained over all the instrument data, rather than that training over the data of individual instrument as shown in the related works \cite{ning2018double,lin2020end}.

\subsubsection{Datasets}
All the compared methods are trained and evaluated with the historical transaction data of the stocks in the China A-shares market. 
The dataset contains (i) minute-level price-volume market information and (ii) the order amount of every trading day for each instrument from Jan. 1, 2017 to June 30, 2019.
Without loss of generality, we focus on intraday order execution, while our method can be easily adapted to a more high-frequency trading scenario.
e{The detailed description of the market information including data preprocessing and order generation can be referred to the supplementary.}

The dataset is divided into training, validation and test datasets according to the trading time.
The detailed statistics are listed in Table~\ref{tab:dataset-statistics}.
We keep only the instruments of CSI 800 in validation and test datasets due to the limit of our computing power.
Note that as CSI 800 Index is designed to reflect the overall performance of A-shares \cite{csi800}, this is a fair setting for evaluating all the compared methods.

For all the model-based methods, the parameters are set based on the training dataset and the performances are directly evaluated on the test dataset. 
For all the learning-based RL methods, the policies are trained on the training dataset and the hyper-parameters are tuned on the validation dataset. 
All these setting values are listed in the supplementary.
The final performances of RL policies we present below are calculated over the test dataset averaging the results from six policies trained with the same hyper-parameter and six different random seeds.

\begin{table}[h]
	\resizebox{\linewidth}{!}{\large
		\begin{tabular}{c|c|c|c}
			\hline
			& Training & Validation  & Test \\
			\hline
			\# instruments & 3,566 & 855 & 855 \\
			\hline
			\# order & 1,654,385 & 35,543 & 33,176\\
			\hline
			Time period & 1/1/2017 - 28/2/2019 & 1/3/2019 - 31/4/2019 & 1/5/2019 - 30/6/2019 \\
			\hline
		\end{tabular}
	}
	\caption{The dataset statistics.}\label{tab:dataset-statistics}
	\vspace{-15pt}
\end{table}

\subsection{Evaluation Workflow}

\begin{figure}[htp]
	\centering
	\includegraphics[width=1.\columnwidth]{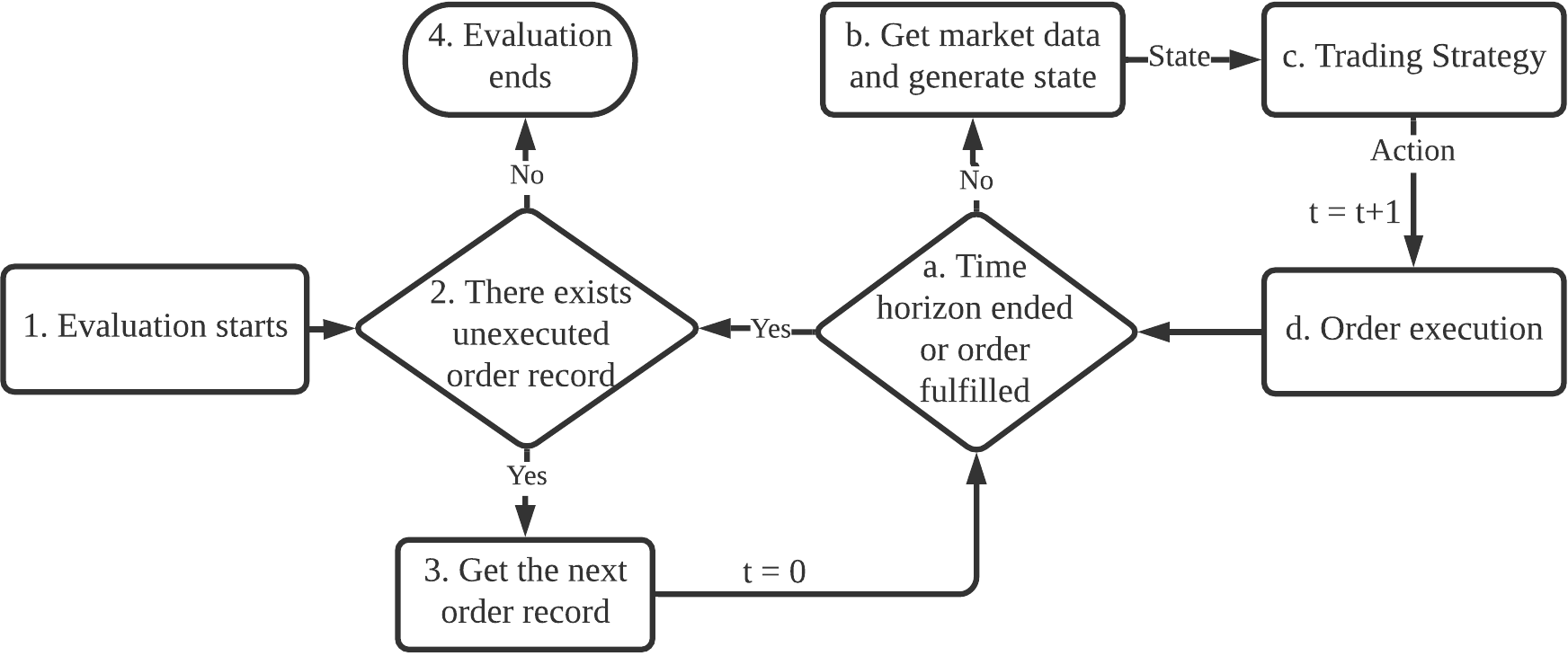}
	\caption{Evaluation workflow.}\label{fig:eva_flow}
	\vspace{-10pt}
\end{figure}

Every trading strategy has been evaluated following the procedure in Figure \ref{fig:eva_flow}.
Given the trading strategy, we need to go through the evaluation data to assess the performance of order execution.
After (1) the evaluation starts, if (2) there still exists unexecuted orders, (3) the next order record would be used for intraday order execution evaluation. Otherwise, (4) the evaluation flow is over.
For each order record, the execution of the order would end once the time horizon ends or the order has been fulfilled. 
After all the orders in the dataset have been executed, the overall performance will be reported. 
The detailed evaluation protocol is presented in supplementary.

\minisection{Evaluation metrics}

We use the obtained \textbf{reward} as the first metric for all the compared methods.
Although some methods are model-based (i.e., TWAP, AC, VWAP) or utilize a different reward function (i.e., PPO), the reward of their trading trajectories would all be calculated as $\frac{1}{|\mathbb{D}|}\sum_{k=1}^{|\mathbb{D}|}\sum_{t=0}^{T-1}R_t^k(\mathbf{s}_t^k, a_t^k)$, where $R_t^k$ is the reward of the $k$-th order at timestep $t$ defined in Eq.~(\ref{eq:instant-reward}) and $|\mathbb{D}|$ is the size of the dataset.
The second is the \textit{price advantage} (\textbf{PA})
$= \frac{10^4}{|\mathbb{D}|}\sum_{k=1}^{|\mathbb{D}|}(\frac{\bar{P}_{\text{strategy}}^k}{\Tilde{p}^k} - 1) ~,$
$\bar{P}_{\text{strategy}}^k$ is the corresponding AEP that our strategy has achieved on order $k$.
This measures the relative gained revenue of a trading strategy compared to that from a baseline price $\Tilde{p}^k$.
Generally, $\Tilde{p}^k$ is the averaged market price of the instrument on the specific trading day, which is constant for all the compared methods.
PA is measured in base points (BPs), where one basis point is equal to 0.01\%.
We conduct significance test \cite{mason2002areas} on PA and reward to verify the statistical significance of the performance improvement.
We also report \textit{gain-loss ratio} (\textbf{GLR}) $ = \frac{\mathbb{E}[\text{PA}|\text{PA}>0]}{\mathbb{E}[\text{PA}|\text{PA}<0]}$ result.

\begin{figure*}[h]
	\centering
	\includegraphics[width=\linewidth]{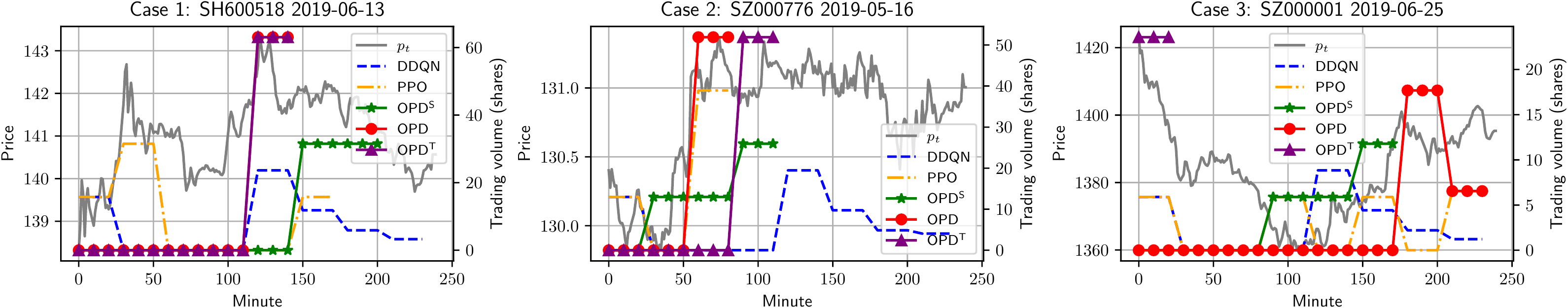}
	\caption{An illustration of execution details of different methods.}\label{fig:case-study}
	\vspace{-10pt}
\end{figure*}

\subsection{Experiment Results}
We analyze the experimental results from three aspects.

\subsubsection{Overall performance}
The overall results are reported in Table~\ref{tab:perf-table} (* indicates p-value $<0.01$ in significance test) and we have the following findings.
(i) \studentg~has the best performance among all the compared methods including \student~without teacher guidance, which illustrates the effectiveness of oracle policy distillation (\textbf{RQ2}).
(ii) All the model-based methods perform worse than the learning-based methods as their adopted market assumptions might not be practical in the real world.
Also, they fail to capture the microstructure of the market and can not adjust their strategy accordingly. 
As TWAP equally distribute the order to every timestep, the AEP of TWAP is always $\Tilde{p}$. 
Thus the PA and GLR results of TWAP are all zero.
(iii) Our proposed method outperforms other learning-based methods (\textbf{RQ1}).
Speaking of DDQN, we may conclude that it is hard for the value-based method to learn a universal Q-function for all instruments.
When comparing \student~with PPO, we conclude that the normalized instant reward function may largely contribute to the performance improvement.

\begin{figure}[t]
	\centering
	\includegraphics[width=0.95\linewidth]{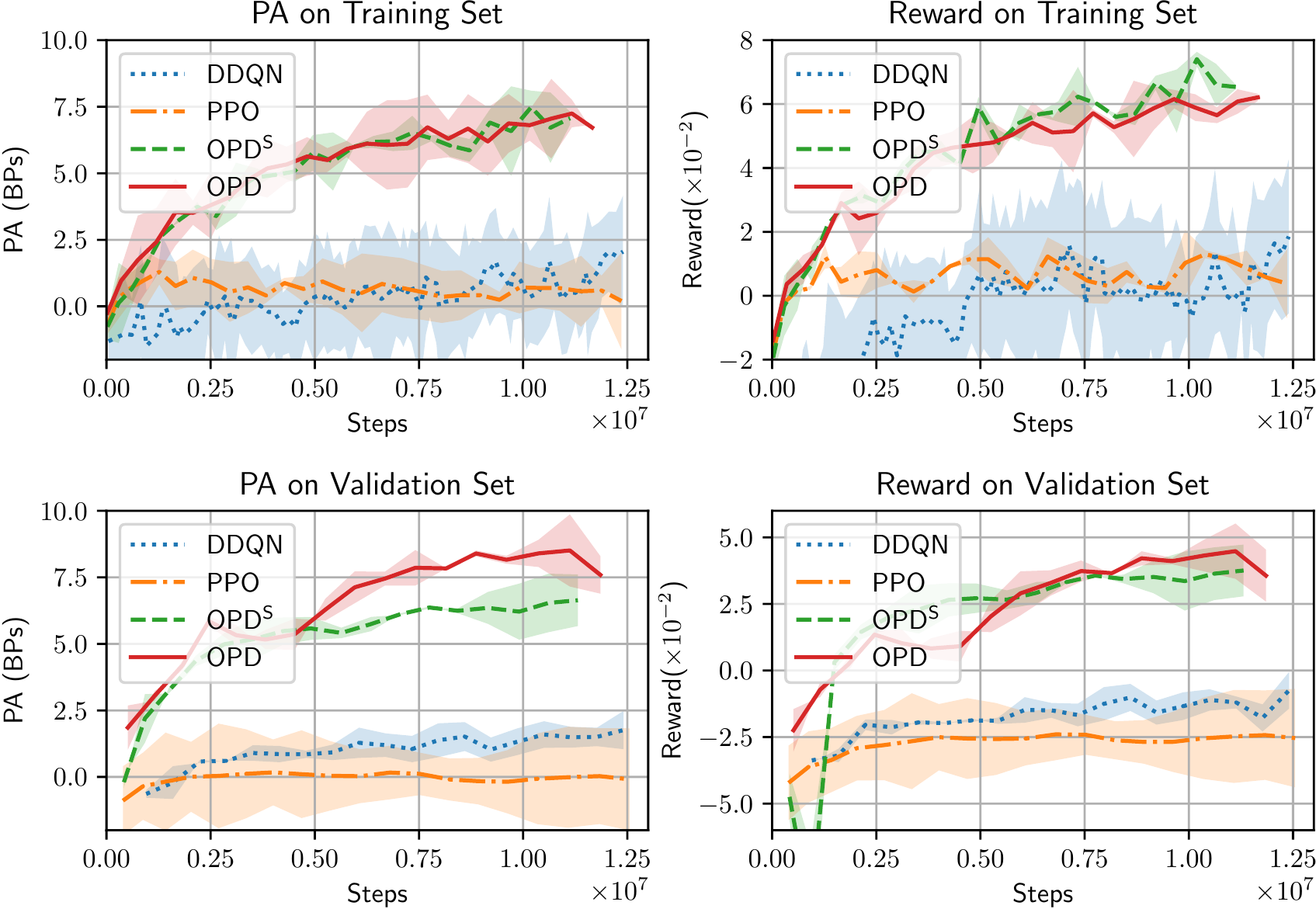}
	\caption{Learning curves (mean$\pm$std over 6 random seeds). Every step means one interaction with the environment.}\label{fig:learning-curve}
	\vspace{-15pt}
\end{figure}

\begin{table}[h]
	\centering
	\resizebox{\columnwidth}{!}{
		\begin{tabular}{c|c|c|c|c}
			\hline
			Category & Strategy &  Reward($\times 10^{-2}$) & PA & GLR \\ \hline
			\multirow{3}*{
            \tabincell{c}{financial\\model-\\based}
			} & TWAP~(Bertsimas et al. 1998)    &          -0.42 &             0 &             0  \\
			 & AC~(Almgren et al. 2001)    &          -1.45 &          2.33 &          0.89  \\ 
			 & VWAP~(Kakade et al. 2004)    &          -0.30 &          0.32 &          0.88  \\ \hline
			\multirow{4}*{
            \tabincell{c}{learning-\\based}
			} & DDQN~(Ning et al. 2018)     &       2.91         &    4.13           &     1.07      \\
			 & PPO~(Lin et al. 2020)    &         1.32 &          2.52 &          0.62  \\ \cline{2-5}
			& \student~(pure student) &          3.24 &          5.19 &          1.19  \\
			& \studentg~(our proposed) & \textbf{3.36*} & \textbf{6.17*} & \textbf{1.35} \\ \hline
		\end{tabular}
	}
	\caption{Performance comparison; the higher, the better.}\label{tab:perf-table}
	\vspace{-10pt}
\end{table}

\subsubsection{Learning analysis}
We show learning curves over training and validation datasets in Figure~\ref{fig:learning-curve}.
We have the following findings that (i) all these methods reach stable convergence after about 8 million steps of training.
(ii) Among all the learning-based methods, our \studentg~method steadily achieve the highest PA and reward results on validation set after convergence, which shows the effectiveness of oracle policy distillation.
(iii) Though the performance gap between \student~and \studentg~is relatively small in the training dataset,
\studentg~evidently outperforms \student~on validation dataset, which shows that oracle guidance can help to learn a more general and effective trading strategy.

\subsubsection{Case study}
We further investigate the action patterns of different strategies (\textbf{RQ3}).
In Figure~\ref{fig:case-study}, we present the execution details of all the learning-based methods. 
The colored lines exhibit the volume of shares traded by these agents at every minute and the grey line shows the trend of the market price $p_t$ of the underlying trading asset on that day.

We can see that in all cases, the \textit{teacher} \teacher~always captures the optimal trading opportunity.
It is reasonable since \teacher~can fully observe the perfect information, including future trends.
The execution situation of DDQN in all cases are exactly the same, which indicates that DDQN fails to adjust its execution strategy dynamically according to the market, thus not performing well.
For PPO and \student, either do they miss the best price for execution or waste their inventory at a bad market opportunity.

Our proposed \studentg~method outperforms all other methods in all three cases except the teacher \teacher.
Specifically, in Case 1, \studentg~captures the best opportunity and manages to find the optimal action sequence.
Although \studentg~fails to capture the optimal trading time in Case 2, it still manages to find a relatively good opportunity and achieve a reasonable and profitable action sequence.
However, as a result of observing no information at the beginning of the trading day, our method tends to miss some opportunities at the beginning of one day, as shown in Case 3.
Though that, in such cases, \studentg~still manages to find a sub-optimal trading opportunity and illustrates some advantages to other methods.

\vspace{-5pt}
\section{Conclusion}
In this paper, we presented a model-free reinforcement learning framework for optimal order execution.
We incorporate a universal policy optimization method that learns and transfers knowledge in the data from various instruments for order execution over all the instruments.
Also, we conducted an oracle policy distillation paradigm to improve the sample efficiency of RL methods and help derive a more reasonable trading strategy.
It has been done through letting a student imitate and distillate the optimal behavior patterns from the optimal trading strategy derived by a teacher with perfect information.
The experiment results over real-world instrument data have reflected the efficiency and effectiveness of the proposed learning algorithm.

In the future, we plan to incorporate policy distillation to learn a general trading strategy from the oracles conducted for each single asset.

\section{Acknowledgments}
The corresponding authors are Kan Ren and Weinan Zhang.
Weinan Zhang is supported by MSRA Joint Research Grant.

\bibliography{ref}

\begin{thebibliography}{39}
\providecommand{\natexlab}[1]{#1}
\providecommand{\url}[1]{\texttt{#1}}
\providecommand{\urlprefix}{URL }
\expandafter\ifx\csname urlstyle\endcsname\relax
  \providecommand{\doi}[1]{doi:\discretionary{}{}{}#1}\else
  \providecommand{\doi}{doi:\discretionary{}{}{}\begingroup
  \urlstyle{rm}\Url}\fi

\bibitem[{Almgren and Chriss(2001)}]{almgren2001optimal}
Almgren, R.; and Chriss, N. 2001.
\newblock Optimal execution of portfolio transactions.
\newblock \emph{Journal of Risk} 3: 5--40.

\bibitem[{Bertsimas and Lo(1998)}]{bertsimas1998optimal}
Bertsimas, D.; and Lo, A.~W. 1998.
\newblock Optimal control of execution costs.
\newblock \emph{Journal of Financial Markets} 1(1): 1--50.

\bibitem[{Bia{\l}kowski, Darolles, and Le~Fol(2008)}]{bialkowski2008improving}
Bia{\l}kowski, J.; Darolles, S.; and Le~Fol, G. 2008.
\newblock Improving VWAP strategies: A dynamic volume approach.
\newblock \emph{Journal of Banking \& Finance} 32(9): 1709--1722.

\bibitem[{Cartea and Jaimungal(2015)}]{cartea2015optimal}
Cartea, A.; and Jaimungal, S. 2015.
\newblock Optimal execution with limit and market orders.
\newblock \emph{Quantitative Finance} 15(8): 1279--1291.

\bibitem[{Cartea and Jaimungal(2016)}]{cartea2016incorporating}
Cartea, A.; and Jaimungal, S. 2016.
\newblock Incorporating order-flow into optimal execution.
\newblock \emph{Mathematics and Financial Economics} 10(3): 339--364.

\bibitem[{Cartea, Jaimungal, and Penalva(2015)}]{cartea2015algorithmic}
Cartea, {\'A}.; Jaimungal, S.; and Penalva, J. 2015.
\newblock \emph{Algorithmic and high-frequency trading}.
\newblock Cambridge University Press.

\bibitem[{Casgrain and Jaimungal(2019)}]{casgrain2019trading}
Casgrain, P.; and Jaimungal, S. 2019.
\newblock Trading algorithms with learning in latent alpha models.
\newblock \emph{Mathematical Finance} 29(3): 735--772.

\bibitem[{Co.(2020)}]{csi800}
Co., C. S.~I. 2020.
\newblock \emph{CSI800 index}.
\newblock \url{https://bit.ly/3btW6di}.

\bibitem[{Dab{\'e}rius, Granat, and Karlsson(2019)}]{daberius2019deep}
Dab{\'e}rius, K.; Granat, E.; and Karlsson, P. 2019.
\newblock Deep Execution-Value and Policy Based Reinforcement Learning for
  Trading and Beating Market Benchmarks.
\newblock \emph{Available at SSRN 3374766} .

\bibitem[{Gaon and Brafman(2020)}]{gaon2020reinforcement}
Gaon, M.; and Brafman, R. 2020.
\newblock Reinforcement Learning with Non-Markovian Rewards.
\newblock In \emph{Proceedings of the AAAI Conference on Artificial
  Intelligence}, volume~34, 3980--3987.

\bibitem[{Green, Vineyard, and Ko{\c{c}}(2019)}]{green2019distillation}
Green, S.; Vineyard, C.~M.; and Ko{\c{c}}, C.~K. 2019.
\newblock Distillation Strategies for Proximal Policy Optimization.
\newblock \emph{arXiv preprint arXiv:1901.08128} .

\bibitem[{Gu{\'e}ant(2016)}]{gueant2016financial}
Gu{\'e}ant, O. 2016.
\newblock \emph{The Financial Mathematics of Market Liquidity: From optimal
  execution to market making}, volume~33.
\newblock CRC Press.

\bibitem[{Gu{\'e}ant and Lehalle(2015)}]{gueant2015general}
Gu{\'e}ant, O.; and Lehalle, C.-A. 2015.
\newblock General intensity shapes in optimal liquidation.
\newblock \emph{Mathematical Finance} 25(3): 457--495.

\bibitem[{Hendricks and Wilcox(2014)}]{hendricks2014reinforcement}
Hendricks, D.; and Wilcox, D. 2014.
\newblock A reinforcement learning extension to the Almgren-Chriss framework
  for optimal trade execution.
\newblock In \emph{2014 IEEE Conference on Computational Intelligence for
  Financial Engineering \& Economics (CIFEr)}, 457--464. IEEE.

\bibitem[{Hinton, Vinyals, and Dean(2015)}]{hinton2015distilling}
Hinton, G.; Vinyals, O.; and Dean, J. 2015.
\newblock Distilling the knowledge in a neural network.
\newblock \emph{arXiv preprint arXiv:1503.02531} .

\bibitem[{Hu(2016)}]{hu2016optimal}
Hu, R. 2016.
\newblock Optimal Order Execution using Stochastic Control and Reinforcement
  Learning.

\bibitem[{Jiang and Liang(2017)}]{jiang2017cryptocurrency}
Jiang, Z.; and Liang, J. 2017.
\newblock Cryptocurrency portfolio management with deep reinforcement learning.
\newblock In \emph{2017 Intelligent Systems Conference (IntelliSys)}, 905--913.
  IEEE.

\bibitem[{Kakade et~al.(2004)Kakade, Kearns, Mansour, and
  Ortiz}]{kakade2004competitive}
Kakade, S.~M.; Kearns, M.; Mansour, Y.; and Ortiz, L.~E. 2004.
\newblock Competitive algorithms for VWAP and limit order trading.
\newblock In \emph{Proceedings of the 5th ACM conference on Electronic
  commerce}, 189--198.

\bibitem[{Lai et~al.(2020)Lai, Zha, Li, and Hu}]{lai2020dual}
Lai, K.-H.; Zha, D.; Li, Y.; and Hu, X. 2020.
\newblock Dual Policy Distillation.
\newblock \emph{arXiv preprint arXiv:2006.04061} .

\bibitem[{Li et~al.(2020)Li, Koyamada, Ye, Liu, Wang, Yang, Zhao, Qin, Liu, and
  Hon}]{li2020suphx}
Li, J.; Koyamada, S.; Ye, Q.; Liu, G.; Wang, C.; Yang, R.; Zhao, L.; Qin, T.;
  Liu, T.-Y.; and Hon, H.-W. 2020.
\newblock Suphx: Mastering Mahjong with Deep Reinforcement Learning.
\newblock \emph{arXiv preprint arXiv:2003.13590} .

\bibitem[{Li et~al.(2019)Li, Zhang, Bian, Tong, and Liu}]{li2019cooperative}
Li, X.; Zhang, J.; Bian, J.; Tong, Y.; and Liu, T.-Y. 2019.
\newblock A cooperative multi-agent reinforcement learning framework for
  resource balancing in complex logistics network.
\newblock In \emph{Proceedings of the 18th International Conference on
  Autonomous Agents and MultiAgent Systems}, 980--988. International Foundation
  for Autonomous Agents and Multiagent Systems.

\bibitem[{Lin and Beling(2020)}]{lin2020end}
Lin, S.; and Beling, P.~A. 2020.
\newblock An End-to-End Optimal Trade Execution Framework based on Proximal
  Policy Optimization.
\newblock In \emph{IJCAI}.

\bibitem[{Mason and Graham(2002)}]{mason2002areas}
Mason, S.~J.; and Graham, N.~E. 2002.
\newblock Areas beneath the relative operating characteristics (ROC) and
  relative operating levels (ROL) curves: Statistical significance and
  interpretation.
\newblock \emph{Quarterly Journal of the Royal Meteorological Society} .

\bibitem[{Mnih et~al.(2015)Mnih, Kavukcuoglu, Silver, Rusu, Veness, Bellemare,
  Graves, Riedmiller, Fidjeland, Ostrovski et~al.}]{mnih2015human}
Mnih, V.; Kavukcuoglu, K.; Silver, D.; Rusu, A.~A.; Veness, J.; Bellemare,
  M.~G.; Graves, A.; Riedmiller, M.; Fidjeland, A.~K.; Ostrovski, G.; et~al.
  2015.
\newblock Human-level control through deep reinforcement learning.
\newblock \emph{Nature} 518(7540): 529--533.

\bibitem[{Moody and Saffell(2001)}]{moody2001learning}
Moody, J.; and Saffell, M. 2001.
\newblock Learning to trade via direct reinforcement.
\newblock \emph{IEEE transactions on neural Networks} 12(4): 875--889.

\bibitem[{Moody and Wu(1997)}]{moody1997optimization}
Moody, J.; and Wu, L. 1997.
\newblock Optimization of trading systems and portfolios.
\newblock In \emph{Proceedings of the IEEE/IAFE 1997 Computational Intelligence
  for Financial Engineering (CIFEr)}, 300--307. IEEE.

\bibitem[{Moody et~al.(1998)Moody, Saffell, Liao, and
  Wu}]{moody1998reinforcement}
Moody, J.~E.; Saffell, M.; Liao, Y.; and Wu, L. 1998.
\newblock Reinforcement Learning for Trading Systems and Portfolios.
\newblock In \emph{KDD}, 279--283.

\bibitem[{Nevmyvaka, Feng, and Kearns(2006)}]{nevmyvaka2006reinforcement}
Nevmyvaka, Y.; Feng, Y.; and Kearns, M. 2006.
\newblock Reinforcement learning for optimized trade execution.
\newblock In \emph{Proceedings of the 23rd international conference on Machine
  learning}, 673--680.

\bibitem[{Ning, Ling, and Jaimungal(2018)}]{ning2018double}
Ning, B.; Ling, F. H.~T.; and Jaimungal, S. 2018.
\newblock Double Deep Q-Learning for Optimal Execution.
\newblock \emph{arXiv preprint arXiv:1812.06600} .

\bibitem[{Parisotto, Ba, and Salakhutdinov(2015)}]{parisotto2015actor}
Parisotto, E.; Ba, J.~L.; and Salakhutdinov, R. 2015.
\newblock Actor-mimic: Deep multitask and transfer reinforcement learning.
\newblock \emph{arXiv preprint arXiv:1511.06342} .

\bibitem[{Rusu et~al.(2015)Rusu, Colmenarejo, Gulcehre, Desjardins,
  Kirkpatrick, Pascanu, Mnih, Kavukcuoglu, and Hadsell}]{rusu2015policy}
Rusu, A.~A.; Colmenarejo, S.~G.; Gulcehre, C.; Desjardins, G.; Kirkpatrick, J.;
  Pascanu, R.; Mnih, V.; Kavukcuoglu, K.; and Hadsell, R. 2015.
\newblock Policy distillation.
\newblock \emph{arXiv preprint arXiv:1511.06295} .

\bibitem[{Schulman et~al.(2017)Schulman, Wolski, Dhariwal, Radford, and
  Klimov}]{schulman2017proximal}
Schulman, J.; Wolski, F.; Dhariwal, P.; Radford, A.; and Klimov, O. 2017.
\newblock Proximal policy optimization algorithms.
\newblock \emph{arXiv preprint arXiv:1707.06347} .

\bibitem[{Silver et~al.(2016)Silver, Huang, Maddison, Guez, Sifre, Van
  Den~Driessche, Schrittwieser, Antonoglou, Panneershelvam, Lanctot
  et~al.}]{silver2016mastering}
Silver, D.; Huang, A.; Maddison, C.~J.; Guez, A.; Sifre, L.; Van Den~Driessche,
  G.; Schrittwieser, J.; Antonoglou, I.; Panneershelvam, V.; Lanctot, M.;
  et~al. 2016.
\newblock Mastering the game of Go with deep neural networks and tree search.
\newblock \emph{nature} 529(7587): 484.

\bibitem[{Sutton and Barto(2018)}]{sutton2018reinforcement}
Sutton, R.~S.; and Barto, A.~G. 2018.
\newblock \emph{Reinforcement learning: An introduction}.
\newblock MIT press.

\bibitem[{Tang et~al.(2019)Tang, Qin, Zhang, Wang, Xu, Ma, Zhu, and
  Ye}]{tang2019deep}
Tang, X.; Qin, Z.; Zhang, F.; Wang, Z.; Xu, Z.; Ma, Y.; Zhu, H.; and Ye, J.
  2019.
\newblock A deep value-network based approach for multi-driver order
  dispatching.
\newblock In \emph{Proceedings of the 25th ACM SIGKDD international conference
  on knowledge discovery \& data mining}, 1780--1790.

\bibitem[{Teh et~al.(2017)Teh, Bapst, Czarnecki, Quan, Kirkpatrick, Hadsell,
  Heess, and Pascanu}]{teh2017distral}
Teh, Y.; Bapst, V.; Czarnecki, W.~M.; Quan, J.; Kirkpatrick, J.; Hadsell, R.;
  Heess, N.; and Pascanu, R. 2017.
\newblock Distral: Robust multitask reinforcement learning.
\newblock In \emph{Advances in Neural Information Processing Systems},
  4496--4506.

\bibitem[{Wu, Gattami, and Flierl(2020)}]{wu2020conditional}
Wu, H.; Gattami, A.; and Flierl, M. 2020.
\newblock Conditional Mutual information-based Contrastive Loss for Financial
  Time Series Forecasting.
\newblock \emph{arXiv preprint arXiv:2002.07638} .

\bibitem[{Ye et~al.(2020)Ye, Pei, Wang, Chen, Zhu, Xiao, Li
  et~al.}]{ye2020reinforcement}
Ye, Y.; Pei, H.; Wang, B.; Chen, P.-Y.; Zhu, Y.; Xiao, J.; Li, B.; et~al. 2020.
\newblock Reinforcement-learning based portfolio management with augmented
  asset movement prediction states.
\newblock \emph{arXiv preprint arXiv:2002.05780} .

\bibitem[{Yin and Pan(2017)}]{yin2017knowledge}
Yin, H.; and Pan, S.~J. 2017.
\newblock Knowledge transfer for deep reinforcement learning with hierarchical
  experience replay.
\newblock In \emph{Thirty-First AAAI Conference on Artificial Intelligence}.

\end{thebibliography}
\end{document}